\documentstyle[11pt,amsmath,amssymb,epsf,epic,cite,epsfig]{article}

\numberwithin{equation}{section}

\newcommand{\tr}{\mbox{Tr}}

\newcommand{\be}{\begin{equation}}
\newcommand{\ee}{\end{equation}}
\newcommand{\bea}{\begin{eqnarray}}
\newcommand{\eea}{\end{eqnarray}}

\newcommand{\e}{{\rm e}}

\renewcommand{\d}{{\rm d}}

\newcommand{\grintl}{[\kern-.18em [}
\newcommand{\grintr}{]\kern-.18em ]}

\newcommand{\rr}{{\rm R}}
\newcommand{\rs}{{\rm S}}
\newcommand{\hh}{{\frak h}}
\newcommand{\av}[1]{\left\langle{#1}\right\rangle}
\newcommand{\rhobar}{{\overline \rho}}
\newcommand{\fd}{{\frak D}}

\newcommand{\usigma}{{\underline{\,\sigma\!}\,}}
\newcommand{\utau}{{\underline{\,\tau\!}\,}}

\newcounter{resultcounter}[section]

\newtheorem{thm}[resultcounter]{Theorem}

\newtheorem{definition}[resultcounter]{Definition}

\def\bed{\begin{definition}}
\def\eed{\end{definition}}

\newcommand{\r}{{\rm R}}
\newcommand{\s}{{\rm S}}
\newcommand{\h}{{\cal H}}
\newcommand{\cx}{{\mathbb C}}
\newcommand{\rx}{{\mathbb R}}
\renewcommand{\i}{{\rm i}}

\newcommand{\fer}[1]{(\ref{#1})}
\newcommand{\scalprod}[2]{\left\langle {#1}, {#2}\right\rangle}
\newcommand{\bbbone}{\mathchoice {\rm 1\mskip-4mu l} {\rm 1\mskip-4mu l}
{\rm 1\mskip-4.5mu l} {\rm 1\mskip-5mu l}}

\begin{document}
\title{Resonant Perturbation
 Theory of Decoherence and Relaxation of Quantum Bits}
\author{M. Merkli\footnote{Department of Mathematics and
Statistics, Memorial University of Newfoundland, St. John's,
 Newfoundland, Canada A1C 5S7. Supported by NSERC
under Discovery Grant 205247. Email: merkli@mun.ca,
http://www.math.mun.ca/$\sim$merkli/ } \and  G.P.
Berman\footnote{Theoretical Division, MS B213, Los Alamos National
Laboratory, Los Alamos, NM 87545, USA. Email: gpb@lanl.gov.  Supported by the
 NNSA of the U. S. DOE at LANL under Contract No. DE-AC52- 06NA25396, and by the IARPA. Publication release number: LA-UR 09-05784.} \and I.M. Sigal\footnote{Department of
Mathematics, University of Toronto, Toronto, Ontario, Canada
M5S2E4. Supported by NSERC under Grant NA 7901.}}

\date{\today}

\maketitle

\begin{abstract}
We  describe our recent results on the resonant perturbation
theory of decoherence and relaxation for quantum system with many
qubits. The approach represents  a rigorous analysis of the
phenomenon of decoherence and relaxation for general $N$-level
systems coupled to reservoirs of the bosonic fields. We derive a
representation of the reduced dynamics valid for all times $t\geq
0$ and for small but fixed interaction strength. Our approach does
not involve master equation approximations and applies to a wide
variety of systems which are not explicitly solvable.
\end{abstract}

\thispagestyle{empty}

\setcounter{page}{1}
\setcounter{section}{1}

\setcounter{section}{0}

\section{Introduction}

Quantum computers (QCs) with large number of quantum bits (qubits)
promise to solve important problems such as factorization of
larger integer numbers, searching large unsorted databases, and
simulations of physical processes exponentially faster than
digital computers. Recently, many efforts were made for designing
scalable (in the number of qubits) QC architectures based on
solid-state implementations. One of the most promising designs of
a solid-state QC is based on superconducting devices with
Josephson junctions and solid-state quantum interference devices
(SQUIDs) serving as qubits (effective spins), which operate in the
quantum regime:  $\hbar\omega >\!\!> k_{\rm B}T$, where $T$ is the
temperature and $\omega$ is the qubit transition frequency. This
condition is widely used in  a superconducting quantum computation
and quantum measurement, when $T\sim 10-20mK$ and $\hbar\omega\sim
100-150mK$ (in temperature units)
\cite{1,2,MSS,YHCCW,DM,Setal,Ketal,Cletal} (see also references
therein). The main advantages of a QC with superconducting qubits
are: (i) The two basic states of a qubit are represented by the
states of a superconducting charge or current in the macroscopic
(several $\mu m$ size) device. The relatively large scale of this
device facilitates the manufacturing, and potential controlling
and measuring of the states of qubits. (ii) The states of charge-
and current-based qubits can be measured using rapidly developing
technologies, such as a single electron transistor, effective
resonant oscillators and micro-cavities with RF amplifiers, and
quantum tunneling effects. (iii) The quantum logic operation can
be implemented exclusively by switching locally on and off
voltages on controlling micro-contacts and magnetic fluxes. (iv)
The devices based on superconducting qubits can potentially
achieve large quantum dephasing and relaxation times of
milliseconds and more at low temperatures, allowing quantum
coherent computation for long enough times. In spite of
significant progress, current devices with superconducting qubits
only have one or two qubits operating with low fidelity even for
simplest operations.

One of the main problems which must be resolved in order to build
a scalable QC is to develop novel approaches for suppression of
unwanted effects such as decoherence and noise. This also requires
to develop the rigorous mathematical tools for analyzing  the
dynamics of decoherence, entanglement and thermalization in order
to control the quantum protocols with needed fidelity. These
theoretical approaches must work for long enough times and be
applicable to both solvable and not explicitly solvable
(non-integrable) systems.

Here we present a review of our results \cite{3,4,5} on the
rigorous analysis of the phenomenon of decoherence and relaxation
for general $N$-level systems coupled to reservoirs. The latter
are described by the bosonic fields.
We suggest a new approach which applies to a wide variety of
systems which are not explicitly solvable. We analyze in detail
the dynamics of an $N$-qubit quantum register collectively coupled
to a thermal environment. Each spin experiences the same
environment interaction, consisting of an energy conserving and an
energy exchange part. We find the decay rates of the reduced
density matrix elements in the energy basis. We show that the
fastest decay rates of off-diagonal matrix elements induced by the
energy conserving interaction is of order $N^2$, while the one
induced by the energy exchange interaction is of the order $N$
only. Moreover, the diagonal matrix elements approach their
limiting values at a rate independent of $N$.
Our method is based on a dynamical quantum resonance theory valid
for small, fixed values of the couplings, and uniformly in time
for $t\geq 0$. We do not make Markov-, Born- or weak coupling (van
Hove) approximations.

\section{Presentation of results}

We consider an $N$-level quantum system $\s$ interacting with a
heat bath $\r$. The former is described by a Hilbert space
$\hh_s=\cx^N$ and a Hamiltonian
 \begin{equation}
H_\s = {\rm diag}(E_1,\ldots,E_N). \label{nlevel}
 \end{equation}
The environment $\rr$ is modelled by a bosonic thermal reservoir
with Hamiltonian
\begin{equation}
H_\rr =\int_{\rx^3} a^*(k) |k| a(k) \d^3k,
\label{5}
\end{equation}
acting on the reservoir Hilbert space $\hh_\r$, and where $a^*(k)$
and $a(k)$ are the usual bosonic creation and annihilation
operators satisfying the canonical commutation relations
$[a(k),a^*(l)]=\delta(k-l)$. It is understood that we consider
$\rr$ in the thermodynamic limit of infinite volume ($\rx^3$) and
fixed temperature $T=1/\beta>0$ (in a phase without condensate).
Given a {\it form factor} $f(k)$, a square integrable function of
$k\in\rx^3$ (momentum representation), the smoothed-out creation
and annihilation operators are defined as $a^*(f) =
\int_{\rx^3}f(k) a^*(k)\d^3k$ and
$a(f)=\int_{\rx^3}\overline{f(k)}a(k)\d^3k$ respectively, and the
hermitian {\it field operator} is
\begin{equation}
\phi(f) = \frac{1}{\sqrt 2}\big[ a^*(f)+a(f)\big].
\label{fieldop}
\end{equation}
The total Hamiltonian, acting on $\hh_\s\otimes\hh_\r$, has the form
\begin{equation}
H=H_\s\otimes\bbbone_\r +\bbbone_\s\otimes H_\r +\lambda v,
\label{01}
\end{equation}
where $\lambda$ is a coupling constant and $v$ is an interaction operator linear in field operators. For simplicity of exposition we consider here initial states where $\s$ and $\r$ are not entangled, and where $\r$ is in thermal equilibrium.\footnote{Our method applies also to initially entangled states, and arbitrary initial states of $\r$ normal w.r.t. the equilibrium state, see \cite{4}.}  The initial density matrix is thus
$$
\rho_0=\rhobar_0\otimes\rho_{\r,\beta},
$$
where $\rhobar_0$ is any state of $\s$ and $\rho_{\r,\beta}$ is the equilibrium state of $\r$ at  temperature $1/\beta$.

Let $A$ be an arbitrary observable of the system (an operator on
the system Hilbert space $\hh_\s$) and set
\begin{equation}
\av{A}_t := \tr_\s(\rhobar_t A) =\tr_{\s +\r}(\rho_t (A\otimes\bbbone_\r) ),
\label{avA}
\end{equation}
where $\rho_t$ is the density matrix of $\s+\r$ at time $t$ and
$$
\rhobar_t={\rm Tr}_{\r}\rho_t
$$
is the reduced density matrix of $\s$. In our approach, the
dynamics of the reduced density matrix $\rhobar_t$ is expressed in
terms of the {\it resonance structure} of the system. Under the
non-interacting dynamics ($\lambda=0$), the evolution of the
reduced density matrix elements of $\s$, expressed in the energy
basis $\{\varphi_k\}_{k=1}^N$ of $H_\s$, is given by
\begin{equation}
[\rhobar_t]_{kl} = \scalprod{\varphi_k}{\e^{-\i
tH_\s}\rhobar_0\e^{\i tH_\s}\varphi_l} = \e^{\i t e_{lk}}
[\rhobar_0]_{kl}, \label{030}
\end{equation}
where $e_{lk}=E_l-E_k$. As the interaction with the reservoir is
turned on, the dynamics \fer{030} undergoes two qualitative
changes.
\begin{itemize}
\item[1.] The ``Bohr frequencies''
\begin{equation}
e\in \{ E-E'\ : E,E'\in{\rm spec}(H_\s)\}
\label{017}
\end{equation}
in the exponent of \fer{030} become {\it complex}, $e\mapsto
\varepsilon_e$. It can be shown generally that the {\it resonance
energies} $\varepsilon_e$ have non-negative imaginary parts,
${\rm Im}\varepsilon_e\geq 0$. If ${\rm Im}\varepsilon_e >0$ then the
corresponding dynamical process is irreversible.
 \item[2.] The matrix elements do not evolve independently any more. Indeed, the
effective energy of $\s$ is changed due to the interaction with
the reservoirs, leading to a dynamics that does not leave
eigenstates of $H_\s$ invariant. (However, to lowest order in the
interaction, the eigenspaces of $H_\s$ are left invariant and
therefore matrix elements with $(m,n)$ belonging to a fixed energy difference $E_m-E_n$ will
evolve in a coupled manner.)
\end{itemize}
Our goal is to derive these two effects from the microscopic
(hamiltonian) model and to quantify them. Our analysis yields the
thermalization and decoherence times of quantum registers.

\subsection{Evolution of reduced dynamics of an $N$-level system}

Let $A\in{\cal B}(\hh_\s)$ be an observable of the system $\s$. We
show in \cite{3,4} that the ergodic averages
\begin{equation}
\av{\av{A}}_\infty := \lim_{T\rightarrow\infty}\frac 1T\int_0^T\av{A}_t \d t
\label{ergav}
\end{equation}
exist, i.e., that $\av{A}_t$ converges in the ergodic sense as
$t\rightarrow\infty$. Furthermore, we show that for any $t\geq 0$
and for any $0<\omega'<2\pi/\beta$,
\begin{equation}
\av{A}_t-\av{\av{A}}_\infty =\sum_{\varepsilon \neq 0}\e^{\i t\varepsilon} R_\varepsilon(A) +O\left(\lambda^2\e^{-[\omega'-O(\lambda)]t}\right),
\label{introdd1}
\end{equation}
where the complex numbers $\varepsilon$ are the  eigenvalues of a
certain explicitly given operator $K(\omega')$, lying in the strip
$\{z\in\cx\ |\ 0\leq {\mathrm Im} z<\omega'/2\}$.  They have the
expansions
\begin{equation}
\varepsilon\equiv \varepsilon_e^{(s)} = e +\lambda^2\delta_e^{(s)}
+ O(\lambda^4), \label{int1}
\end{equation}
where $e\in{\mathrm spec}(H_\s\otimes\bbbone_\s-\bbbone_\s\otimes
H_\s) = {\mathrm spec}(H_\s)- {\mathrm spec}(H_\s)$ and the
$\delta_e^{(s)}$ are the eigenvalues of a matrix $\Lambda_e$,
called a {\it level-shift operator}, acting on the eigenspace of
$H_\s\otimes\bbbone_\s-\bbbone_\s\otimes H_\s$ corresponding to
the eigenvalue $e$ (which is a subspace of $\hh_\s\otimes\hh_\s$).
The  $R_\varepsilon(A)$ in \fer{introdd1} are linear functionals
of $A$ and are given in terms of the initial state, $\rho_0$, and
certain operators depending on the Hamiltonian $H$. They have the
expansion
\begin{equation}
R_\varepsilon(A)=\sum_{(m,n)\in I_e}\varkappa_{m,n} A_{m,n} +O(\lambda^2),
\label{p02}
\end{equation}
where $I_e$ is the collection of all pairs of indices such that $e=E_m-E_n$, the $E_k$ being the eigenvalues of $H_\s$. Here, $A_{m,n}$ is the $(m,n)$-matrix element of the observable $A$ in the energy basis of $H_\s$, and the $\varkappa_{m,n}$ are coefficients depending on the initial state of the system (and on $e$, but not on $A$ nor on $\lambda$).

\smallskip
{\bf Discussion.\ } -- In the absence of interaction
($\lambda=0$) we have $\varepsilon=e\in{\mathbb R}$, see \fer{int1}.
Depending on the interaction each resonance energy $\varepsilon$ may
migrate into the upper complex plane, or it may stay on the real
axis, as $\lambda\neq 0$.

-- The averages $\av{A}_t$ approach their ergodic means
$\av{\av{A}}_\infty$ if and only if ${\rm Im} \varepsilon
>0$ for all $\varepsilon\neq 0$. In this case the convergence takes place
on the time scale $[{\mathrm Im}\varepsilon]^{-1}$. Otherwise
$\av{A}_t$ oscillates. A sufficient condition for decay is that
${\rm Im}\delta_e^{(s)} > 0$ (and $\lambda$ small, see
\fer{int1}).

-- The error term in \fer{introdd1} is small in $\lambda$,
uniformly in $t\geq 0$, and it decays in time quicker than any of
the main terms in the sum on the r.h.s.: indeed, ${\rm
Im}\varepsilon =O(\lambda^2)$ while $\omega'-O(\lambda)>\omega'/2$
independent of small values of $\lambda$. However, this means that
we are in the regime $\lambda^2<\!\!<\omega'<2\pi/\beta$ (see
before \fer{introdd1}), which implies that $\lambda^2$ must be
much smaller than the temperature $T=1/\beta$. Using a more
refined analysis one can get rid of this condition, see also
remarks on p.376 of \cite{4}.

-- Relation \fer{p02} shows that to lowest order in the
perturbation the group of (energy basis) matrix elements of any
observable $A$ corresponding to a fixed energy difference
$E_m-E_n$ evolve jointly, while those of different such groups
evolve independently.

\medskip
It is well known that there are two kinds of processes which drive
decay (or irreversibility) of $\s$: energy-exchange processes
characterized by $[v,H_\s]\neq 0$ and energy preserving ones
 where $[v,H_\s]=0$. The former are induced by interactions having
nonvanishing probabilities for processes of absorption and
emission of field quanta with energies corresponding to the Bohr
frequencies of $\s$ and thus typically inducing thermalization of
$\s$. Energy preserving interactions suppress such processes,
allowing only for a phase change of the system during the
evolution (``phase damping'',
\cite{PSE,BKT,DBV,DG,JZKGKS,MP,SGC}).

To our knowledge, energy-exchange systems have so far been treated
using Born and Markov master equation approximations (Lindblad
form of dynamics) or they have been studied numerically, while for
energy conserving systems one often can find an exact solution.
The present representation \fer{introdd1} gives a detailed picture
of the dynamics of averages of observables for interactions with
and without energy exchange. The resonance energies $\varepsilon$
and the functionals $R_\varepsilon$ can be calculated for concrete
models, as illustrated in the next section. We mention that the
resonance dynamics representation can be used to study the
dynamics of entanglement of qubits coupled to local and collective
reservoirs. Work on this is in progress.

\medskip
{\bf Contrast with weak coupling approximation.\ } Our
representation \fer{introdd1} of the true dynamics of $\s$ relies
only on the smallness of the coupling parameter $\lambda$, and no
approximation is made. In the absence of an exact solution, it is
common to make a weak coupling Lindblad  master equation
approximation of the dynamics, in which the reduced density matrix
evolves according to $\rhobar_t=\e^{t {\cal L}}\rhobar_0$, where
$\cal L$ is the {\it Lindblad generator}, \cite{BP,Da1,Da2}. This
approximation can lead to results that differ qualitatively from
the true behaviour. For instance, the Lindblad master equation
predicts that the system $\s$ approaches its Gibbs state at the
temperature of the reservoir in the limit of large times. However,
it is clear that in reality, the {\it coupled} system $\s+\rr$
will approach equilibrium, and hence the asymptotic state of $\s$
alone, being the reduction of the coupled equilibrium state, is
the Gibbs state of $\s$ only to first approximation in the
coupling (see also illustration below, and references \cite{3,4}).
In particular, the system's asymptotic density matrix is not
diagonal in the original energy basis, but it has off-diagonal
matrix elements of $O(\lambda^2)$.
Features of this kind cannot be captured by the Lindblad
approximation, but are captured in our approach.

It has been shown (see e.g. \cite{DF,Da1,Da2,JP}) that the weak
coupling limit dynamics generated by the Lindblad operator is
obtained in the regime $\lambda\rightarrow 0$,
$t\rightarrow\infty$, with $t/\lambda^2$ fixed. One of the
strengths of our approach is that we do not impose any relation
between $\lambda$ and $t$, and our results are valid for all times
$t\geq 0$, provided $\lambda$ is small. It has been observed
\cite{DF,JP} that for certain systems of the type $\s+\rr$, the
second order contribution of the exponents $\varepsilon_e^{(s)}$
in \fer{int1} correspond to eigenvalues of the Lindblad generator.
Our resonance method gives the true exponents, i.e., we do not
lose the contributions of any order in the interaction. If the
energy spectrum of $H_\s$ is degenerate, it happens that the
second order contributions to ${\rm Im} \varepsilon_e^{(s)}$
vanish. This corresponds to a Lindblad generator having several
real eigenvalues. In this situation the correct dynamics (approach
to a final state) can be captured only by taking into account
higher order contributions to the exponents $\varepsilon_e^{(s)}$,
see \cite{M}. To our knowledge, so far this can only be done with
the method presented in this paper, and is beyond the reach of the
weak coupling method.

\bigskip
{\bf Illustration: single qubit.\ } Consider $\s$ to be a single
spin $1/2$ with energy gap $\Delta=E_2-E_1>0$. $\s$ is coupled to
the heat bath $\r$ via the operator
\begin{equation}
v = \left[
\begin{array}{cc}
 a & c\\
\overline c & b
\end{array}
\right] \otimes\phi(g), \label{michael*}
\end{equation}
where $\phi(g)$ is the Bose field operator \fer{fieldop}, smeared
out with a coupling function (form factor) $g(k)$, $k\in\rx^3$,
and the $2\times 2$ coupling matrix (representing the coupling
operator in the energy eigenbasis) is hermitian. The operator
\fer{michael*} - or a sum of such terms, for which our technique
works equally well - is the most general coupling which is linear
in field operators. We refer to \cite{4} for a discussion of the
link between \fer{michael*} and the spin-boson model. We take $\s$
initially in a coherent superposition in the energy basis,
\begin{equation}
\rhobar_0=\textstyle \frac 12 \left[
\begin{array}{cc}
 1 & 1\\
 1 & 1
\end{array}
\right]. \label{illust}
\end{equation}
In \cite{4} we derive from representation \fer{introdd1} the
following expressions for the dynamics of matrix elements,
for all $t\geq 0$:
\begin{eqnarray}
[\rhobar_t]_{m,m} &=& \frac{\e^{-\beta E_m}}{Z_{\s,\beta}} +
\frac{(-1)^m}{2} \tanh\left(\frac{\beta\Delta}{2}\right)
\e^{\i t\varepsilon_0(\lambda)} +R_{m,m}(\lambda,t) ,\ \ \ m=1,2, \ \
\label{nn2}\\
{} [\rhobar_t]_{1,2} &=& {\textstyle \frac 12} \e^{\i
t\varepsilon_{-\Delta}(\lambda)} +R_{1,2}(\lambda,t),\label{nn3}
\end{eqnarray}
where the resonance energies $\varepsilon$ are given by
\begin{eqnarray}
\varepsilon_0(\lambda) &=& \i\lambda^2 \pi^2|c|^2\xi(\Delta) +O(\lambda^4)\nonumber\\
\varepsilon_\Delta(\lambda) &=& \Delta +\lambda^2 R +{\textstyle
\frac \i2}\lambda^2\pi^2\left[ |c|^2 \xi
(\Delta)+(b-a)^2\xi(0)\right]  +O(\lambda^4)\label{i4}\\
\varepsilon_{-\Delta}(\lambda) &=&
-\overline{\varepsilon_\Delta(\lambda)}\nonumber
\end{eqnarray}
with
\begin{equation}
\xi(\eta) = \lim_{\epsilon\downarrow 0} \frac
1\pi\int_{\rx^3}\d^3k\coth\!\left(\frac{\beta |k|}{2}\right)
 |g(k)|^2 \frac{\epsilon}{(|k|-\eta)^2+\epsilon^2},
\label{35}
\end{equation}
and
$$
R = {\textstyle \frac 12}(b^2-a^2) \scalprod{g}{\omega^{-1}g}
$$
$$+{\textstyle \frac 12}|c|^2 {\rm P.V.}\int_{\rx\times S^2} u^2|g(|u|,\sigma)|^2
\coth\!\left(\frac{\beta |u|}{2}\right)\frac{1}{u-\Delta}.$$ The
remainder terms in \fer{nn2}, \fer{nn3} satisfy
$|R_{m,n}(\lambda,t)|\leq C\lambda^2$, uniformly in $t\geq 0$, and
they can be decomposed into a sum of a constant and a decaying
part,
$$
R_{m,n}(\lambda,t) = \av{\av{p_{n,m}}}_\infty-\delta_{m,n}
\frac{\e^{-\beta E_m}}{Z_{\s,\beta}} +R'_{m,n}(\lambda,t),
$$
where $|R'_{m,n}(\lambda,t)|=O(\lambda^2\e^{-\gamma t})$, with
$\gamma=\min\{{\rm Im}\varepsilon_0, {\rm
Im}\varepsilon_{\pm\Delta}\}$. These relations show that

-- To second order in $\lambda$, convergence of the populations to
the equilibrium values (Gibbs law), and decoherence occur
exponentially fast, with rates $\tau_T=[{\mathrm
Im}\varepsilon_0(\lambda)]^{-1}$ and $\tau_D=[{\mathrm
Im}\varepsilon_\Delta(\lambda)]^{-1}$, respectively. (If either of
these imaginary parts vanishes then the corresponding process does
not take place, of course.) In particular, coherence of the
initial state stays preserved on time scales of the order
$\lambda^{-2}[|c|^2 \xi(\Delta)+(b-a)^2\xi(0)]^{-1}$, c.f.
\fer{i4}.

-- The final density matrix of the spin is {\it not} the Gibbs
state of the qubit, and it is {\it not} diagonal in the energy
basis. The deviation of the final state from the Gibbs state is
given by $\lim_{t\rightarrow\infty}R_{m,n}(\lambda,t) =
O(\lambda^2)$. This is clear heuristically too, since typically
the {\it entire} system $\s+\r$ approaches its joint equilibrium
in which $\s$ and $\r$ are entangled. The reduction of this state
to $\s$ is the Gibbs state of $\s$ modulo $O(\lambda^2)$ terms
representing a shift in the effective energy of $\s$ due to the
interaction with the bath. In this sense, coherence in the energy
basis of $\s$ is created by thermalization. We have quantified
this in \cite{4}, Theorem 3.3.

-- In a markovian master equation approach the above phenomenon
(i.e., variations of $O(\lambda^2)$ in the time-asymptotic limit)
cannot be detected. Indeed in that approach one would conclude
that $\s$ approaches its Gibbs state as $t\rightarrow\infty$.

\subsection{Collective decoherence of a qubit register}
\label{sectcolldec}

 In the sequel we analyze in more detail the evolution of a
qubit register of size $N$. The Hamiltonian is
\begin{equation}
H_\s = \sum_{i,j=1}^N J_{ij}S^z_i S^z_j +\sum_{j=1}^N B_j S_j^z,
\label{02}
\end{equation}
where the $J_{ij}$ are pair interaction constants and $B_j$ is the value of a magnetic field at the location of spin $j$. The Pauli spin operator is
\begin{equation}
S^z= \left[
\begin{array}{cc}
1 & 0\\
0 & -1
\end{array}
\right]
\label{04}
\end{equation}
and $S_j^z$ is the matrix $S^z$ acting on the $j$-th spin only.

We consider a {\it collective coupling}  between the register $\s$ and the reservoir $\rr$: the distance between the $N$ qubits is much smaller
than the correlation length of the reservoir and as a consequence, each qubit
feels {\it the same} interaction with the reservoir. The corresponding interaction operator is (compare with \fer{01})
\begin{equation}
\lambda_1 v_1+\lambda_2 v_2 = \lambda_1\sum_{j=1}^N S_j^z\otimes
\phi(g_1) + \lambda_2\sum_{j=1}^N S_j^x\otimes\phi(g_2).
 \label{06}
\end{equation}
Here $g_1$ and $g_2$ are form factors and the coupling constants $\lambda_1$ and $\lambda_2$ measure the strengths of the {\it energy conserving} (position-position) coupling, and the {\it energy exchange} (spin flip) coupling, respectively. Spin-flips are implemented by the $S_j^x$ in \fer{06}, representing the Pauli matrix
\begin{equation}
S^x =\ \left[
\begin{array}{cc}
0 & 1\\
1 & 0
\end{array}
\right]
 \label{08}
\end{equation}
acting on the $j$-th spin.
The total Hamiltonian takes the form \fer{01} with $\lambda v$ replaced by \fer{06}. It is convenient to represent $\rhobar_t$ as a matrix in the energy basis, consisting of
eigenvectors $\varphi_\usigma$ of $H_\s$. These are vectors in $\hh_\s = \cx^2\otimes\cdots \otimes\cx^2 = \cx^{2^N}$ indexed by spin configurations
\begin{equation}
\usigma=\{\sigma_1,\ldots,\sigma_N\}\in \{+1,-1\}^N,\qquad
\varphi_\usigma =
\varphi_{\sigma_1}\otimes\cdots\otimes\varphi_{\sigma_N},
\label{011}
\end{equation}
where
\begin{equation}
\varphi_+=\left[
\begin{array}{c}
1\\
0
\end{array}
\right],\quad
 \varphi_-=\left[
\begin{array}{c}
0\\
1
\end{array}
\right],
\label{012}
\end{equation}
so that
\begin{equation}
H_\s\varphi_\usigma = E(\usigma)\varphi_\usigma \mbox{\qquad with
\qquad} E(\usigma) = \sum_{i,j=1}^N J_{ij}\sigma_i\sigma_j
+\sum_{j=1}^N B_j\sigma_j.
\label{013}
\end{equation}
 We denote the reduced
density matrix elements as
\begin{equation}
[\rhobar_t]_{\usigma,\utau} =
\scalprod{\varphi_\usigma}{\rhobar_t\varphi_\utau}.
\label{014}
\end{equation}
The Bohr frequencies \fer{017} are now
\begin{equation}
e(\usigma,\utau) = E(\usigma)-E(\utau)= \sum_{i,j=1}^N J_{ij}(\sigma_i\sigma_j-\tau_i\tau_j) +\sum_{j=1}^N B_j(\sigma_j-\tau_j)
 \label{e}
\end{equation}
and they become complex resonance energies $\varepsilon_e=
\varepsilon_e(\lambda_1,\lambda_2)\in\cx$ under perturbation.

\medskip
{\it Assumption of non-overlapping resonances.\ } The Bohr
frequencies \fer{e} represent ``unperturbed'' energy levels and we
follow their motion under perturbation ($\lambda_1, \lambda_2$).
In this work, we consider the regime of non-overlapping
resonances, meaning that the interaction is small relative to the
spacing of the Bohr frequencies.

\smallskip
 We
show in \cite{4}, Theorem 2.1, that for all $t\geq 0$,
\begin{eqnarray}
[\rhobar_t]_{\usigma,\utau} -
\langle\!\langle[\rhobar_\infty]_{\usigma,\utau} \rangle\!\rangle &=& \sum_{\{e:\ \varepsilon_e\neq 0\}} \e^{\i t\varepsilon_e}  \Big[  \sum_{\usigma',\utau'} w^{\varepsilon_e}_{\usigma,\utau;\usigma',\utau'}\ [\rhobar_0]_{\usigma',\utau'} +O(\lambda_1^2+\lambda_2^2)\Big]\nonumber \\
&&  +
O\big((\lambda_1^2+\lambda_2^2)\e^{-[\omega'+O(\lambda)]t}\big).
\label{intro1}
\end{eqnarray}
This result is obtained by specializing \fer{introdd1} to the
specific system at hand and considering observables
$A=|\varphi_\utau\rangle\langle\varphi_\usigma|$. In \fer{intro1},
we have in accordance with \fer{ergav}
$\langle\!\langle[\rhobar_\infty]_{\usigma,\utau} \rangle\!\rangle
=\lim_{T\rightarrow\infty} \frac 1T \int_0^T
[\rhobar_t]_{\usigma,\utau} \ \d t$. The coefficients $w$ are
overlaps of resonance eigenstates which vanish unless
$e=-e(\usigma,\utau)=-e(\usigma',\utau')$ (see point 2. after
\fer{017}). They represent the dominant contribution to the
functionals $R_\varepsilon$ in \fer{introdd1}, see also \fer{p02}.
The $\varepsilon_e$ have the expansion
\begin{equation}
\varepsilon_e\equiv \varepsilon^{(s)}_e = e +\delta_e^{(s)}
+O(\lambda_1^4+\lambda_2^4), \label{2.2}
\end{equation}
where the label $s=1,\ldots,\nu(e)$ indexes the splitting of the
eigenavlue $e$ into $\nu(e)$ distinct resonance energies. The
lowest order corrections $\delta_e^{(s)}$ satisfy
\begin{equation}
\delta_e^{(s)}=O(\lambda_1^2+\lambda_2^2 ).
\label{intro2}
\end{equation}
They are the (complex) eigenvalues of an operator $\Lambda_e$,
called the {\it level shift operator} associated to $e$. This
operator acts on the eigenspace of $L_\s$ associated to the
eigenvalue $e$ (a subspace of the qubit register Hilbert space;
see \cite{4,5} for the formal definition of $\Lambda_e$). It
governs the lowest order shift of eigenvalues under perturbation.
One can see by direct calculation that ${\rm Im}\delta_e^{(s)}\geq
0$.

\medskip

{\bf Discussion.\ } -- To lowest order in the perturbation, the
group of reduced density matrix elements
$[\rhobar_t]_{\usigma,\utau}$ associated to a fixed
$e=e(\usigma,\utau)$ evolve in a coupled way, while groups of
matrix elements associated to different $e$ evolve independently.

-- The density matrix elements of a given group mix and evolve in
time according to the weight functions $w$ and the exponentials
$\e^{\i t\varepsilon_e^{(s)}}$. In the absence of interaction
($\lambda_1=\lambda_2=0$) all the $\varepsilon_e^{(s)}=e$ are
real. As the interaction is switched on, the $\varepsilon_e^{(s)}$
typically migrate into the upper complex plane, but they may stay
on the real line (due to some symmetry or due to an `inefficient
coupling').

-- The matrix elements $[\rhobar_t]_{\usigma,\utau}$ of a group
$e$ approach their ergodic means if and only if all the nonzero
$\varepsilon_e^{(s)}$ have strictly positive imaginary part. In
this case the convergence takes place on a time scale of the order
$1/\gamma_e$, where
\begin{equation}
\gamma_e=\min\left\{ {\rm Im}\varepsilon_e^{(s)}:\
s=1,\ldots,\nu(e) {\ \rm s.t.\ } \varepsilon_e^{(s)}\neq 0\right\}
 \label{3.17.1}
\end{equation}
is the decay rate of the group associated to $e$. If an
$\varepsilon_e^{(s)}$ stays real then the matrix elements of the
corresponding group oscillate in time. A sufficient condition for
decay of the group associated to $e$ is $\gamma_e>0$, i.e. ${\rm
Im}\delta_e^{(s)}>0$ for all $s$, and $\lambda_1$, $\lambda_2$
small.

\medskip {\bf Decoherence rates.\ } We illustrate our results on
decoherence rates for a qubit register with $J_{ij}=0$ (the
general case is treated in \cite{5}).  We consider {\it generic}
magnetic fields defined as follows. For $n_j\in\{0,\pm 1,\pm2\}$,
$j=1,\ldots,N$, we have
\begin{equation}
\sum_{j=1}^N B_jn_j=0\qquad \Longleftrightarrow \qquad n_j=0\
\forall j. \label{3.1}
\end{equation}
Condition \fer{3.1} is satisfied generically in the sense that
only for very special choices of $B_j$ does it not hold (one such
special choice is $B_j={\mbox constant}$). For instance, if the
$B_j$ are chosen to be independent, and uniformly random from an
interval $[B_{\rm min},B_{\rm max}]$, then \fer{3.1} is satisfied
with probability one. We show in \cite{5}, Theorem 2.3, that the
decoherence rates \fer{3.17.1} are given by
\begin{equation}
\gamma_e = \left\{
\begin{array}{ll}
\lambda_1^2 y_1(e)+\lambda_2^2 y_2(e) +y_{12}(e)
, & e\neq 0\\
\lambda_2^2 y_0, & e=0
\end{array}
\right\} + O(\lambda_1^4+\lambda_2^4). \label{3.22.1}
\end{equation}
Here, $y_1$ is a contributions coming from the energy conserving
interaction, $y_0$ and $y_2$ are due to the spin flip interaction.
The term $y_{12}$ is due to both interactions and is of
$O(\lambda^2_1+\lambda^2_2)$. We give explicit expressions for
$y_0$, $y_1$, $y_2$ and $y_{12}$ in \cite{5}, Section 2. For the
present purpose, we limit ourselves to discussing the properties
of the latter quantities.

\begin{itemize}

\item[-] {\it Properties of $y_1(e)$:} $y_1(e)$ vanishes if either
$e$ is such that $e_0:=\sum_{j=1}^n(\sigma_j-\tau_j)=0$, or the
infra-red behaviour of the coupling function $g_1$ is too regular
(in three dimensions $g_1\propto |k|^p$ with $p>-1/2$). Otherwise
$y_1(e)>0$. Moreover, $y_1(e)$ is proportional to the temperature
$T$.

\item[-] {\it Properties of $y_2(e)$:} $y_2(e)>0$ if
$g_2(2B_j,\Sigma)\neq 0$ for all $B_j$ (form factor
$g_2(k)=g_2(|k|,\Sigma)$ in spherical coordinates). For low
temperatures $T$, $y_2(e)\propto T$, for high temperatures
$y_2(e)$ approaches a constant.

\item[-] {\it Properties of $y_{12}(e)$:} If either of
$\lambda_1$, $\lambda_2$ or $e_0$ vanish, or if $g_1$ is infra-red
regular as mentioned above, then $y_{12}(e)= 0$. Otherwise
$y_{12}(e)>0$, in which case $y_{12}(e)$ approaches constant
values for both $T\rightarrow 0, \infty$.

\item[-] {\it Full decoherence:} If $\gamma_e>0$ for all $e\neq 0$
then all off-diagonal matrix elements approach their limiting
values exponentially fast. In this case we say that full
decoherence occurs. It follows from the above points that we have
full decoherence if $\lambda_2\neq 0$ and $g_2(2B_j,\Sigma)\neq 0$
for all $j$, and provided $\lambda_1,\lambda_2$ are small enough
(so that the remainder term in \fer{3.22.1} is small). Note that
if $\lambda_2=0$ then matrix elements associated to energy
differences $e$ such that $e_0=0$ will not decay on the time scale
given by the second order in the perturbation ($\lambda_1^2$). \\
We point out that generically, $\s+\rr$ will reach a joint
equilibrium as $t\rightarrow\infty$, which means that the final
reduced density matrix of $\s$ is its Gibbs state modulo a
peturbation of the order of the interaction between $\s$ and
$\rr$, see \cite{3,4}. Hence generically, the density matrix of
$\s$ does not become diagonal in the energy basis as
$t\rightarrow\infty$.

\item[-] {\it Properties of $y_0$:} $y_0$ depends on the energy
exchange interaction only. This reflects the fact that for a
purely energy conserving interaction, the populations are
conserved \cite{3,4,PSE}. If $g_2(2B_j,\Sigma)\neq 0$ for all $j$,
then $y_0>0$ (this is sometimes called the ``Fermi Golden Rule
Condition''). For small temperatures $T$, $y_0\propto T$, while
$y_0$ approaches a finite limit as $T\rightarrow\infty$.
\end{itemize}

In terms of complexity analysis, it is important to discuss the
{\it dependence of $\gamma_e$ on the register size $N$}.

\begin{itemize}
\item[-] We show in \cite{5} that $y_0$ is independent of $N$.
This means that the thermalization time, or relaxation time of the
diagonal matrix elements (corresponding to $e=0$), is $O(1)$ in
$N$.

\item[-] To determine the order of magnitude of the decay rates of
the off-diagonal density matrix elements (corresponding to $e\neq
0$) relative to the register size $N$, we assume the magnetic
field to have a certain distribution denoted by $\av{\ }$. We show
in \cite{5} that
\begin{equation}
\av{y_1}=y_1\propto e_0^2,\quad  \av{y_2}=C_B{\frak
D}(\usigma-\utau), \quad \mbox{and}\quad \av{y_{12}}=
c_B(\lambda_1,\lambda_2) N_0(e), \label{3.20'}
\end{equation}
where $C_B$ and $c_B=c_B(\lambda_1,\lambda_2)$ are positive
constants (independent of $N$), with $c_B(\lambda_1,\lambda_2)=
O(\lambda_1^2+\lambda_2^2)$. Here, $N_0(e)$ is the number of
indices $j$ such that $\sigma_j=\tau_j$ for each $(\usigma,\utau)$
s.t. $e(\usigma,\utau)=e$, and
\begin{equation}
\fd (\usigma-\utau):= \sum_{j=1}^N |\sigma_j-\tau_j|
 \label{3.20}
\end{equation}
is the {\it Hamming distance} between the spin configurations
$\usigma$ and $\utau$ (which depends on $e$ only).

\item[-] Consider $e\neq 0$. It follows from
\fer{3.22.1}-\fer{3.20} that for purely energy conserving
interactions ($\lambda_2=0$), $\gamma_e\propto \lambda_1^2
e_0^2=\lambda_1^2 [\sum_{j=1}^N(\sigma_j-\tau_j)]^2$, which can be
as large as $O(\lambda_1^2 N^2)$. On the other hand, for purely
energy exchanging interactions ($\lambda_1=0$), we have
$\gamma_e\propto \lambda_2^2 \fd(\usigma-\utau)$, which cannot
exceed $O(\lambda_2^2 N)$. If both interactions are acting, then
we have the additional term $\av{y_{12}}$, which is of order
$O((\lambda_1^2+\lambda_2^2)N)$. This shows the following:

{\it The fastest decay rate of reduced off-diagonal density matrix
elements due to the energy conserving interaction alone is of
order $\lambda_1^2 N^2$, while the fastest decay rate due to the
energy exchange interaction alone is of the order $\lambda_2^2 N$.
Moreover, the decay of the diagonal matrix elements is of oder
$\lambda_1^2$, i.e., independent of $N$.}

\end{itemize}

\bigskip

\noindent {\bf Remarks.\ } 1. For $\lambda_2=0$ the model can be
solved explicitly \cite{PSE}, and one shows that the fastest
decaying matrix elements have decay rate proportional to
$\lambda_1^2 N^2$. Furthermore, the model with a {\it
non-collective, energy-conserving interaction}, where each qubit
is coupled to an independent reservoir, can also be solved
explicitly \cite{PSE}. The fastest decay rate in this case is
shown to be proportional to $\lambda_1^2 N$.

2. As mentioned at the beginning of this section, we take the
coupling constants $\lambda_1$, $\lambda_2$ so small that the
resonances do not overlap. Consequently $\lambda_1^2 N^2$ and
$\lambda_2^2N$ are bounded above by a constant proportional to the
gradient of the magnetic field in the present situation, see also
\cite{5}. Thus the decay rates $\gamma_e$ do not increase
indefinitely with increasing $N$ in the regime considered here.
Rather, the $\gamma_e$ are attenuated by small coupling constants
for large $N$. They are of the order $\gamma_e\sim \Delta$.  We
have shown that modulo an overall, common ($N$-dependent)
prefactor, the decay rates originating from the energy conserving
and exchanging interactions differ by a factor $N$.

3. Collective decoherence has been studied extensively in the
literature. Among the many theoretical, numerical and experimental
works we mention here only \cite{AHWBK,BKT,DG,FF,PSE}, which are
closest to the present work. We are not aware of any prior work
giving explicit decoherence rates of a register for not explicitly
solvable models, and without making master equation technique
approximations.

\section{Resonance representation of reduced dynamics}

The goal of this section is to give a precise statement of the
core representation \fer{introdd1}, and to outline the main ideas
 behind the proof of it.

The $N$-level system is coupled to the reservoir (see also
\fer{nlevel}, \fer{5}) through the operator
\begin{equation}
v=\sum_{r=1}^R \lambda_r G_r\otimes\phi(g_r),
\label{vop}
\end{equation}
where each $G_r$ is a hermitian $N\times N$ matrix, the $g_r(k)$
are form factors and the $\lambda_r\in\rx$ are coupling constants.
Fix any phase $\chi\in\rx$ and define
\begin{equation}
g_{r,\beta}(u,\sigma) := \sqrt{\frac{u}{1-\e^{-\beta u}}}\ |u|^{1/2}
\left\{
\begin{array}{ll}
g_r(u,\sigma) & \mbox{if $u\geq 0$},\\
-\e^{\i\chi} \overline g_r(-u,\sigma) & \mbox{if $u<0$},
\end{array}
\right.
\label{7}
\end{equation}
where $u\in\rx$ and $\sigma \in S^2$. The phase $\chi$ is a
parameter which can be chosen appropriately as to satisfy the
following condition.

\medskip
{\bf (A)} The map $\omega\mapsto g_{r,\beta}(u+\omega,\sigma)$ has
an analytic extension to a complex neighbourhood $\{|z|<\omega'\}$
of the origin, as a map from $\cx$ to $L^2(\rx^3,\d^3k)$.

\medskip
Examples of $g$ satisfying (A) are given by  $g(r,\sigma) = r^p
\e^{-r^m} g_1(\sigma)$, where $p=-1/2+n$, $n=0,1,\ldots$, $m=1,2$,
and $g_1(\sigma)=\e^{\i\phi}\overline g_1(\sigma)$.

This condition ensures that the technically simplest version of
the dynamical resonance theory, based on complex spectral
translations, can be implemented. The technical simplicity comes
at a price: on the one hand, it limits the class of admissible
functions $g(k)$, which have to behave appropriately in the
infra-red regime so that the parts of \fer{7} fit nicely together
at $u=0$, to allow for an analytic continuation. On the other
hand, the square root in \fer{7} must be analytic as well, which
implies the condition $\omega'<2\pi/\beta$.

It is convenient to introduce the doubled Hilbert space ${\cal
H}_\s=\hh_\s\otimes\hh_\s$, whose normalized vectors accommodate
any state on the system $\s$ (pure or mixed). The trace state, or
infinite temperature state, is represented by the vector
\begin{equation}
\Omega_\s = \frac{1}{\sqrt N}\sum_{j=1}^N\varphi_j\otimes\varphi_j
\label{p1}
\end{equation}
via
\begin{equation}
{\cal B}(\hh_\s)\ni A\mapsto
\scalprod{\Omega_\s}{(A\otimes\bbbone) \Omega_\s}. \label{p2}
\end{equation}
Here the $\varphi_j$ are the orthonormal eigenvectors of $H_\s$.
This is just the Gelfand-Naimark-Segal construction for the trace
state. Similarly, let ${\cal H}_\r$ and $\Omega_{\rr,\beta}$ be
the Hilbert space and the vector representing the equilibrium
state of the reservoirs at inverse temperature $\beta$. In the
{\it Araki-Woods} representation of the field, we have ${\cal
H}_\r={\cal F}\otimes{\cal F}$, where $\cal F$ is the bosonic Fock
space over the one-particle space $L^2(\rx^3,\d^3k)$ and
$\Omega_{\rr,\beta}=\Omega\otimes\Omega$, $\Omega$ being the Fock
vacuum of $\cal F$ (see also \cite{4,5} for more detail). Let
$\psi_0\otimes\Omega_{\r,\beta}$ be the vector in
$\h_\s\otimes\h_\r$ representing the density matrix at time $t=0$.
It is not difficult to construct the unique operator in
$B\in\bbbone_\s\otimes{\hh_\s}$ satisfying
$$
B\Omega_\s=\psi_0.
$$
(See also \cite{4} for concrete examples.) We define the reference
vector
$$
\Omega_{\rm ref} := \Omega_\s\otimes\Omega_{\r,\beta}
$$
and set
$$
\lambda=\max_{r=1,\ldots,R}|\lambda_r|.
$$
\begin{thm}[Dynamical resonance theory \cite{3,4,5}]
\label{prop3.2} Assume condition (A) with a fixed $\omega'$
satisfying $0<\omega'<2\pi/\beta$. There is a constant $c_0$ s.t.
if $\lambda\leq c_0/\beta$ then the limit $\av{\av{A}}_\infty$,
\fer{ergav}, exists for all observables $A\in{\cal B}(\hh_\s)$.
Moreover, for all such $A$ and for all $t\geq 0$ we have
\begin{eqnarray}
\av{A}_t - \av{\av{A}}_\infty &=&\sum_{e,s:\
\varepsilon_e^{(s)}\neq 0 } \ \ \sum_{s=1}^{\nu(e)}\e^{\i
t\varepsilon_e^{(s)}}
 \scalprod{(B^*\psi_0)\otimes \Omega_{\r,\beta}}{Q_e^{(s)}
 (A\otimes\bbbone_\s) \Omega_{\rm ref}}\nonumber \\
&& +O(\lambda^2\e^{-[\omega'+O(\lambda)] t}). \label{24}
\end{eqnarray}
The $\varepsilon_e^{(s)}$ are given by \fer{int1}, $1\leq \nu(e)
\leq {\rm mult}(e)$ counts the splitting of the eigenvalue $e$
into distinct resonance energies $\varepsilon_e^{(s)}$ and the
$Q_e^{(s)}$ are (non-orthogonal) finite-rank projections.
\end{thm}

\medskip This result is the basis for a detailed analysis of the
reduced dynamics of concrete systems, like the $N$-qubit register
introduced in Section \ref{sectcolldec}. We obtain \fer{intro1}
(in particular, the overlap functions $w$) from \fer{24} by
analyzing the projections $Q_e^{(s)}$ in more detail. Let us
explain how to link the overlap $\scalprod{(B^*\psi_0)\otimes
\Omega_{\r,\beta}}{Q_e^{(s)} (A\otimes\bbbone_\s) \Omega_{\rm
ref}}$ to its initial value for a non-degenerate Bohr energy $e$,
and where $A=|\varphi_n\rangle\langle\varphi_m|$. (The latter
observables used in \fer{introdd1} give the matrix elements of the
reduced density matrix in the energy basis.)

The $Q_e^{(s)}$ is the spectral (Riesz) projection of an operator
$K_\lambda$ associated with the eigenvalue $\varepsilon_e^{(s)}$,
see \fer{eq23}.\footnote{In reality, we consider a {\it spectral
deformation} $K_\lambda(\omega)$, where $\omega$ is a complex
parameter. This is a technical trick to perform our analysis.
Physical quantities do not depend on $\omega$ and therefore we do
not display this parameter here.}  If a Bohr energy $e$,
\fer{017}, is simple, then there is a single resonance energy
$\varepsilon_e$ bifurcating out of $e$, as $\lambda\neq 0$. In
this case the projection $Q_e\equiv Q_e^{(s)}$ has rank one,
$Q_e=|\chi_e\rangle\langle\widetilde\chi_e|$, where $\chi_e$ and
$\widetilde\chi_e$ are eigenvectors of $K_\lambda$ and its
adjoint, with eigenvalue $\varepsilon_e$ and its complex
conjugate, respectively, and
$\scalprod{\chi_e}{\widetilde\chi_e}=1$. From perturbation theory
we obtain
$\chi_e=\widetilde\chi_e=\varphi_k\otimes\varphi_l\otimes\Omega_{\rr,\beta}+
O(\lambda)$, where $H_\s\varphi_j=E_j\varphi_j$ and $E_k-E_l=e$.
The overlap in the sum of \fer{24} becomes
\begin{eqnarray}
\lefteqn{ \scalprod{(B^*\psi_0)\otimes \Omega_{\r,\beta}}{Q_e
(A\otimes\bbbone_\s) \Omega_{\rm ref} }}\nonumber\\
&=& \scalprod{(B^*\psi_0)\otimes\Omega_{\rr,\beta}}{
|\varphi_k\otimes\varphi_l\otimes\Omega_{\r,\beta}\rangle\langle
\varphi_k\otimes\varphi_l\otimes\Omega_{\r,\beta}|
(A\otimes\bbbone_\s) \Omega_{\rm ref}} +O(\lambda^2)\nonumber\\
&=&\scalprod{B^*\psi_0}{|\varphi_k\otimes\varphi_l
\rangle\langle\varphi_k\otimes\varphi_l | (A\otimes
\bbbone_\s)\Omega_\s}+O(\lambda^2)
 \label{n1}
\end{eqnarray}
The choice $A=|\varphi_n\rangle\langle\varphi_m|$ in \fer{avA}
gives $\av{A}_t =[\rhobar_t]_{m,n}$, the reduced density matrix
element. With this choice of $A$, the main term in \fer{n1}
becomes (see also \fer{p1})
\begin{eqnarray}
\lefteqn{
 \scalprod{B^*\psi_0}{|\varphi_k\otimes\varphi_l
\rangle\langle\varphi_k\otimes\varphi_l | (A\otimes
\bbbone_\s)\Omega_\s} }\nonumber\\
&=&\frac{1}{\sqrt N}\delta_{kn}\delta_{lm}\scalprod{
B^*\psi_0}{\varphi_n\otimes\varphi_m}\nonumber\\
&=&\delta_{kn}\delta_{lm}\scalprod{
B^*\psi_0}{(|\varphi_n\rangle\langle \varphi_m|\otimes
\bbbone_\s)\Omega_\s}\nonumber\\
&=&\delta_{kn}\delta_{lm}\scalprod{
\psi_0}{(|\varphi_n\rangle\langle \varphi_m|\otimes
\bbbone_\s)B\Omega_\s}\nonumber\\
&=&\delta_{kn}\delta_{lm}[\rhobar_0]_{mn}. \label{n2}
\end{eqnarray}
In the second-last step, we commute $B$ to the right through
$|\varphi_n\rangle\langle \varphi_m|\otimes \bbbone_\s$, since $B$
belongs to the commutant of the algebra of observables of $\s$. In
the last step, we use $B\Omega_\s=\psi_0$.

Combining \fer{n1} and \fer{n2} with Theorem \ref{prop3.2} we
obtain, in case $e=E_m-E_n$ is a simple eigenvalue,
\begin{eqnarray*}
[\rhobar_t]_{mn} - \langle\!\langle[\rhobar_\infty]_{mn}
\rangle\!\rangle = \sum_{\{e,s: \varepsilon_e^{(s)}\neq 0\}}\e^{\i
t\varepsilon_e^{(s)}} \big[ \delta_{kn}\delta_{lm}
[\rhobar_0]_{mn} +O(\lambda^2)\big]
+O(\lambda^2\e^{-[\omega'+O(\lambda)] t}).\
\end{eqnarray*}
This explains the form \fer{intro1} for a simple Bohr energy $e$.
The case of degenerate $e$ (i.e., where several different pairs of
indices $k,l$ satisfy $E_k-E_l=e$) is analyzed along the same
lines, see \cite{5} for details.

\subsection{Mechanism of dynamical resonance theory,
 outline of proof of Theorem \ref{prop3.2}}

Consider any observable $A\in B(\hh_\rs)$. We have
\begin{eqnarray}
\langle A\rangle_t &=& {\rm Tr}_\rs\left[\rhobar_t \ A \right]\nonumber\\
&=& {\rm Tr}_{\rs+\rr}\left[ \rho_t \ A \otimes\bbbone_\rr\right]\nonumber\\
&=&\scalprod{\psi_0}{\e^{\i tL_\lambda}\left[
A\otimes\bbbone_\rs\otimes \bbbone_\rr\right]\e^{-\i
tL_\lambda}\psi_0}. \label{2}
\end{eqnarray}
In the last step, we pass to the {\it representation Hilbert
space} of the system (the GNS Hilbert space), where the initial
density matrix is represented by the vector $\psi_0$ (in
particular, the Hilbert space of the small system becomes ${\frak
h}_\s \otimes {\frak h}_\s$), see also before equations \fer{p1},
\fer{p2}. As mentioned above, in this review we consider initial
states where $\s$ and $\r$ are not entangled. The initial state is
represented by the product vector
$\psi_0=\Omega_\s\otimes\Omega_{\rr,\beta}$, where $\Omega_{\rs}$
is the trace state of $\rs$, \fer{p2}, $\scalprod{\Omega_{\rs}}{
(A\otimes\bbbone_\rs)\Omega_{\rs}} = \frac{1}{N}{\rm Tr\,}(A)$,
and where $\Omega_{\rr,\beta}$ is the equilibrium state of $\rr$
at a fixed inverse temperature $0<\beta<\infty$. The dynamics is
implemented by the group of automorphisms $\e^{\i
tL_\lambda}\cdot\e^{-\i tL_\lambda}$. The self-adjoint generator
$L_\lambda$ is called the {\it Liouville operator}. It is of the
form $L_\lambda = L_0+\lambda W$, where $L_0=L_\rs+L_\rr$
represents the uncoupled Liouville operator, and $\lambda W$ is
the interaction \fer{vop} represented in the GNS Hilbert space. We
refer to \cite{4,5} for the specific form of $W$.

We borrow a trick from the analysis of open systems far from
equilibrium: there is a (non-self-adjoint) generator $K_\lambda$
s.t.
\begin{eqnarray*}
\e^{\i tL_\lambda} A \e^{-\i tL_\lambda} &=& \e^{\i tK_\lambda} A
\e^{-\i tK_\lambda}
\mbox{\ \ for all observables $A$, $t\geq 0$, and}\\
K_\lambda\psi_0 &=& 0.
\end{eqnarray*}
$K_\lambda$ can be constructed in a standard way, given
$L_\lambda$ and the reference vector $\psi_0$. $K_\lambda$ is of
the form $K_\lambda =L_0+\lambda I$, where the interaction term
undergoes a certain modification ($W\rightarrow I$), c.f.
\cite{4}. As a consequence, formally, we may replace the
propagators in \fer{2} by those involving $K$. The resulting
propagator which is directly applied to $\psi_0$ will then just
disappear due to the invariance of $\psi_0$. One can carry out
this procedure in a rigorous manner, obtaining the following
resolvent representation \cite{4}
\begin{equation}
\langle A\rangle_t = -\frac{1}{2\pi \i}\int_{{\mathbb R}-\i}
\scalprod{\psi_0}{(K_\lambda(\omega)-z)^{-1}
\left[A\otimes\bbbone_\rs\otimes\bbbone_\rr\right]\psi_0} \e^{\i
tz}\d z, \label{eq20}
\end{equation}
where $K_\lambda(\omega)=L_0(\omega)+\lambda I(\omega)$, $I$ is
representing the interaction, and $\omega\mapsto
K_\lambda(\omega)$ is a spectral deformation (translation) of
$K_\lambda$. The latter is constructed as follows. There is a
deformation transformation $U(\omega)=\e^{-\i\omega D}$, where $D$
is the (explicit) self-adjoint generator of translations
\cite{4,5,MMS} transforming the operator $K_\lambda$ as
\begin{equation}
K_\lambda(\omega) = U(\omega) K_\lambda U(\omega)^{-1} =
L_0+\omega N +\lambda I(\omega). \label{eq18}
\end{equation}

\begin{figure}[h]
\centerline{
\includegraphics[width=11cm]{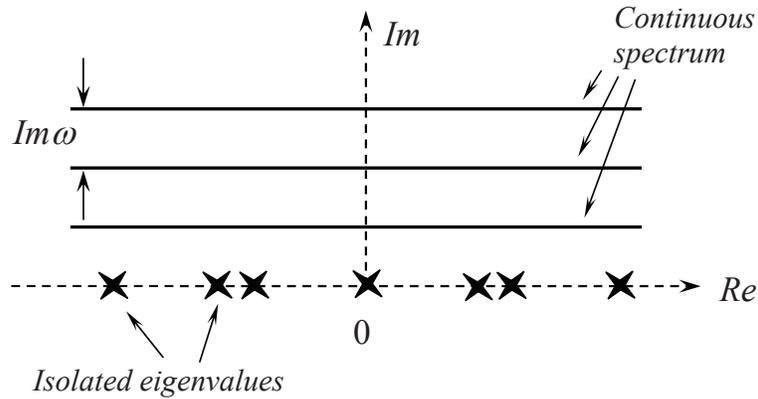}}
 \caption{Spectrum of $K_0(\omega)$}
\end{figure}

 Here, $N=N_1\otimes\bbbone +\bbbone\otimes N_1$ is the total
number operator of a product of two bosonic Fock spaces ${\cal
F}\otimes{\cal F}$ (the Gelfand-Naimark-Segal Hilbert space of the
reservoir), and where $N_1$ is the usual number operator on
$\mathcal F$. $N$ has spectrum ${\mathbb N}\cup\{0\}$, where $0$
is a simple eigenvalue (with vacuum eigenvector
$\Omega_{\rr,\beta}=\Omega\otimes\Omega$). For real values of
$\omega$, $U(\omega)$ is a group of unitaries. The spectrum of
$K_\lambda(\omega)$ depends on ${\rm Im\,}\omega$ and moves
according to the value of ${\rm Im\,}\omega$, whence the name
``spectral deformation''. Even though $U(\omega)$ becomes
unbounded for complex $\omega$, the r.h.s. of \fer{eq18} is a well
defined closed operator on a dense domain, analytic in $\omega$ at
zero. Analyticity is used in the derivation of \fer{eq20} and this
is where the analyticity condition (A) after \fer{7} comes into
play. The operator $I(\omega)$ is infinitesimally small with
respect to the number operator $N$. Hence we use perturbation
theory in $\lambda$ to examine the spectrum of
$K_\lambda(\omega)$.

\begin{figure*}[t]
\centering
\includegraphics[width=10cm]{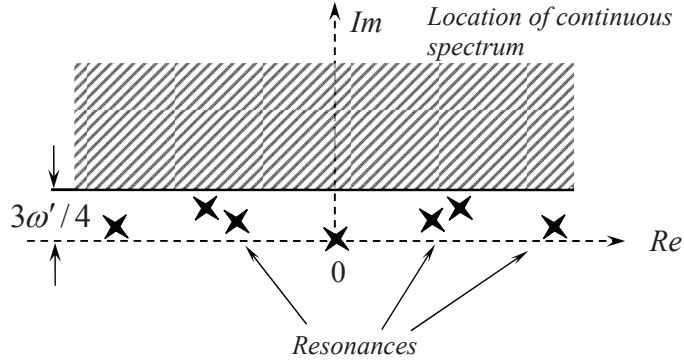}
\vspace*{-.5cm} \caption{Spectrum of $K_\lambda(\omega)$.
 Resonances $\varepsilon_e^{(s)}$ are uncovered.}
\end{figure*}

The point of the spectral deformation is that the (important part
of the) spectrum of $K_\lambda(\omega)$ is much easier to analyze
than that of $K_\lambda$, because the deformation uncovers the
resonances of $K_\lambda$. We have (see Figure 1)
$$
{\rm
spec\,}\big(K_{0}(\omega)\big)=\{E_i-E_j\}_{i,j=1,\ldots,N}\bigcup_{n\geq
1}\{\omega n +{\mathbb R}\},
$$
because $K_0(\omega)=L_0+\omega N$, $L_0$ and $N$ commute, and the
eigenvectors of $L_0=L_\rs+L_\rr$ are
$\varphi_i\otimes\varphi_j\otimes\Omega_{\rr,\beta}$. Here, we have
$H_\s\varphi_j = E_j\varphi_j$. The continuous spectrum is bounded
away from the isolated eigenvalues by a gap of size ${\rm
Im\,}\omega$. For values of the coupling parameter $\lambda$ small
compared to ${\rm Im\,}\omega$, we can follow the displacements of
the eigenvalues by using analytic perturbation theory. (Note that
for ${\rm Im\,}\omega=0$, the eigenvalues are imbedded into the
continuous spectrum, and analytic perturbation theory is not
valid! The spectral deformation is indeed very useful!)

\begin{thm}[\cite{4}]
\label{thm3} (See Fig. 2.) Fix ${\rm Im\,}\omega$ s.t. $0<{\rm
Im\,}\omega<\omega'$ (where $\omega'$ is as in Condition (A)).
There is a constant $c_0>0$ s.t. if $|\lambda|\leq c_0/\beta$
then, for all $\omega$ with ${\rm Im\,}\omega>7\omega'/8$, the
spectrum of $K_\lambda(\omega)$ in the complex half-plane $\{{\rm
Im\,}z<\omega'/2\}$ is independent of $\omega$ and consists purely
of the distinct eigenvalues
$$
\{\varepsilon_e^{(s)}\ :\  e\in{\rm spec}(L_\rs),
s=1,\ldots,\nu(e)\},
$$
where $1\leq\nu(e)\leq {\rm mult}(e)$ counts the splitting of the
eigenvalue $e$. Moreover,
$$\lim_{\lambda\rightarrow
0}|\varepsilon_e^{(s)}(\lambda)-e|=0
$$
for all $s$, and we have
${\rm Im\,}\varepsilon_e^{(s)}\geq 0$. Also, the continuous
spectrum of $K_\lambda(\omega)$ lies in the region $\{{\rm
Im\,}z\geq 3\omega'/4\}$.
\end{thm}

Next we separate the contributions to the path integral in
\fer{eq20} coming from the singularities at the resonance energies
and from the continuous spectrum. We deform the path of
integration $z={\mathbb R}-\i$ into the line $z={\mathbb R}+\i
\omega'/2$, thereby picking up the residues of poles of the
integrand at $\varepsilon_e^{(s)}$ (all $e$, $s$). Let ${\mathcal
C}_e^{(s)}$ be a small circle around $\varepsilon_e^{(s)}$, not
enclosing or touching any other spectrum of $K_\lambda(\omega)$.
We introduce the generally non-orthogonal Riesz spectral
projections
\begin{equation}
Q_e^{(s)}  = Q_e^{(s)}(\omega,\lambda) = -\frac{1}{2\pi\i}
\int_{{\mathcal C}_e^{(s)}} (K_\lambda(\omega)-z)^{-1} \d z.
\label{eq23}
\end{equation}

\begin{figure}[h]
\centerline{
\includegraphics[width=11cm]{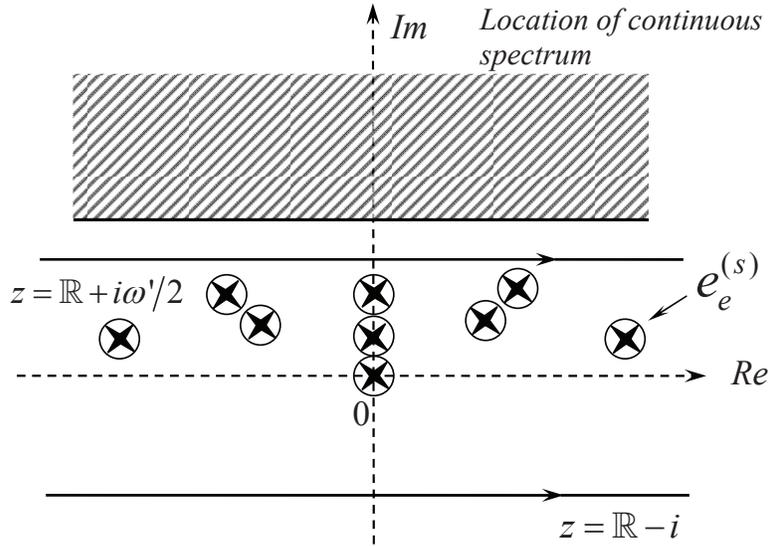}}
\vspace*{-.5cm} \caption{Contour deformation: $\int_{{\mathbb
R}-\i}\d z =  \sum_{e,s}\int_{{\mathcal C}_e^{(s)}}\d z
+\int_{{\mathbb R}+\i\omega'/2} \d z$ }
\end{figure}

It follows from \fer{eq20} that
\begin{equation}
\langle A\rangle_t = \sum_e\sum_{s=1}^{\nu(e)} \e^{\i
t\varepsilon^{(s)}_e} \scalprod{\psi_0}{Q_e^{(s)}
[A\otimes\bbbone_\rs\otimes\bbbone_\rr]\psi_0}
+O(\lambda^2\e^{-\omega' t/2}). \label{eq21}
\end{equation}
Note that the imaginary parts of all resonance energies
$\varepsilon_e^{(s)}$ are smaller than $\omega'/2$, so that the
remainder term in \fer{eq21} is not only small in $\lambda$, but
it also decays faster than all of the terms in the sum. (See also
Figure 3.) We point out also that instead of deforming the path
integration contour as explained before \fer{eq23}, we could
choose $z=\rx+\i[\omega'-O(\lambda)]$, hence transforming the
error term in \fer{eq21} into the one given in \fer{24}.

Finally, we notice that all terms in \fer{eq21} with
$\varepsilon_e^{(s)}\neq 0$ will vanish in the ergodic mean limit,
so
$$
\langle\!\langle A\rangle\!\rangle_\infty
=\lim_{T\rightarrow\infty}\frac 1T \int_0^T \langle A\rangle_t \
\d t =\sum_{s: \varepsilon_0^{(s)} =0}\scalprod{\psi_0}{Q_0^{(s)}
[A\otimes\bbbone_\rr\otimes\bbbone_\rr]\psi_0}.
$$
We now see that the linear functionals \fer{p02} are represented
as
\begin{equation}
R_{\varepsilon_e^{(s)}}(A)= \scalprod{\psi_0}{Q_e^{(s)}
[A\otimes\bbbone_\rs\otimes\bbbone_\rr]\psi_0}. \label{eq22}
\end{equation}
This concludes the outline of the proof of Theorem \ref{prop3.2}.


\begin{thebibliography}{}



\bibitem{AHWBK}
Altepeter, J.B., Hadley, P.G., Wendelken, S.M., Berglund, A.J.,
Kwiat, P.G.: {\it Experimental Investigation of a Two-Qubit
Decoherence-Free Subspace.} Phys. Rev. Lett. {\bf 92}, no.14,
147901



\bibitem{BP}
Breuer, H.-P., Petruccione, F., {\it The theory of open quantum
systems.} Oxford University Press 2002


\bibitem{BKT}
Berman, G.P., Kamenev, D.I., Tsifrinovich, V.I.: {\it Collective
decoherence of the superpositional entangled states in the quantum
Shor algorithm}. Phys. Rev. A {\bf 71}, 032346 (2005)





\bibitem{Cletal}
Clerk, A.A., Devoret, M.H., Girvin, S.M., Marquardt, F., and
Schoelkopf, R.J.: {\it  Introduction to quantum noise, measurement and
amplifcation.}  Preprint arXiv:0810.4729v1 [cond-mat] (2008).



\bibitem{Da1}
Davies, E.B.: {\it Markovian master equations.} Comm. Math. Phys.
{\bf 39}, 91 (1974)


\bibitem{Da2}
Davies, E.B.: {\it Markovian Master Equations. II} Math. Ann. {\bf
219}, 147-158 (1976)



\bibitem{DBV}
Dalvit, D.A.R., Berman, G.P., Vishik, M.: {\it Dynamics of bosonic quantum systems in coherent state
representation}. Phys. Rev. A, {\bf 73}, 013803 (2006)


\bibitem{DM}
Devoret, M.H. and Martinis, J.M.: {\it Implementing qubits with
superconducting integrated circuits.} Quantum Information
Processing {\bf 3}, 163-203 (2004).


\bibitem{DF}
Derezi\'nski, J., Fr\"uboes, R.: {\it Fermi Golden Rule and Open
Quantum Systems}, in Lecture Notes in Mathematics 1882, Springer
Verlag 2006


\bibitem{DG}
Duan, L.-M., Guo, G.-C.: {\it Reducing decoherence in quantum-computer memory with all quantum bits coupling to the same environment}. Phys. Rev. A {\bf 57}, no.2, 737-741 (1998)




\bibitem{FF}
Fedorov, A., Fedichkin, L.: {\it Collective decoherence of nuclear
spin clusters.} J. Phys. Condens. Matter {\bf 18}, 3217-3228
(2006)



\bibitem{JP}

Jak\u si\'c, V., Pillet, C.-A.: {\it From resonances to master
equations} Ann. Inst. Henri Poincar\'e {\bf 67}, no.4, 425-445
(1997)


\bibitem{JZKGKS}
Joos, E., Zeh, H.D., Kiefer, C., Giulini, D., Kupsch, J. Stamatescu, I.O.: Decoherence and the appearence of a classical world in quantum theory. Second edition. {\it Springer Verlag, Berlin, 2003}





\bibitem{Ketal}
Katz, N.,  Neeley, M., Ansmann, M., Bialczak, R.C.,  Hofheinz, M., Lucero, E.,  O'Connell, A.: {\it  Reversal of the weak measurement of a
quantum state in a superconducting phase qubit}. Phys. Rev.
Lett. {\bf  101}, 200401-1-4 (2008).




\bibitem{2}
 Kinion, D. and Clarke, J.: {\it Microstrip superconducting quantum interference device radio-frequency amplifier: Scattering parameters and input coupling.} Appl. Phys. Lett. {\bf 92}, 172503 (2008)





\bibitem{MSS}
Makhlin, Y., Sch\"on, G., and Shnirman, A.: {\it  Quantum-state
engineering with Josephson-junction devices.}  Rev. Mod.
Phys.{\bf 73}, 380-400 (2001).



\bibitem{M}
Merkli, M.: {\it Level shift operators for open quantum systems.}
J. Math. Anal. Appl.  {\bf 327},  no. 1, 376--399  (2007)





\bibitem{3}
Merkli, M.,  Sigal, I.M., and G.P. Berman, G.P.: {\it  Decoherence and Thermalization.} Phys. Rev. Lett. {\bf 98}, 130401 (2007).



\bibitem{4}
Merkli, M.,  Sigal, I.M., and Berman, G.P.: {\it  Resonance theory of
decoherence and thermalization.} Annals of Physics {\bf 323}, 373
(2008).

\bibitem{5}
 Merkli, M.,  Berman, B.P., and Sigal, I.M. :{\it  Dynamics of collective decoherence and thermalization.}  Annals of Physics {\bf 323}, 3091 (2008).



\bibitem{MMS}
Merkli, M., M\"uck, M., Sigal, I.M.: {\it Instability of
Equilibrium States for Coupled Heat Reservoirs at Different
Temperatures.} J. Funct. Anal.  {\bf 243} no. 1, 87-120 (2007)




\bibitem{MP}
Mozyrsky, D., Privman, V.: {\it Adiabatic Decoherence}. Journ.
Stat. Phys. {\bf 91}, 3/4, 787-799 (1998)







\bibitem{PSE}
Palma, G.M., Suominen, K.-A., Ekert, A.K.: Quantum Computers and
Dissipation. {\em Proc. R. Soc. Lond. A} {\bf 452}, 567-584 (1996)




\bibitem{SGC}
Shao, J., Ge, M.-L., Cheng, H.: {\it Decoherence of quantum-nondemolition systems}. Phys. Rev. E, {\bf 53}, no.1, 1243 - 1245 (1996)




\bibitem{Setal}
Steffen, M.,  Ansmann, M., McDermott, R., Katz, N., Bialczak, R.C.,
Lucero, E., Neeley, M., Weig, E.M., Cleland, A.N. and Martinis, J.M.: {\it
State tomography of capacitively shunted phase qubits with high
fidelity.} Phys. Rev. Lett. {\bf  97}  050502-1-4 (2006).



\bibitem{1}
Wendin, G., and Shumeiko, V. S.: {\it Quantum bits with Josephson
junctions.} (Review Article), Low Temp. Phys. {\bf 33}, 724 (2007).



\bibitem{YHCCW}
Yu, Y.,  Han, S., Chu, X.,  Chu, S.I.,  Wang, Z.: {\it Coherent temporal
oscillations of macroscopic quantum states in a Josephson
junction}, SCIENCE  {\bf 296}, 889-892 (2002).




\end{thebibliography}
\end{document}